\pdfoutput=1 
\documentclass[aps,pra,reprint,nofootinbib]{revtex4-1} 
\usepackage[utf8]{inputenc}
\usepackage[margin=1in]{geometry}
\usepackage{amsmath,amsfonts,amssymb}
\usepackage{xcolor}
\definecolor{darkblue}{rgb}{0, 0, 0.8}
\usepackage{physics}
\usepackage{graphicx}
\usepackage{hyperref}
\hypersetup{
	colorlinks   = true, 
	urlcolor     = darkblue,
	linkcolor    = darkblue, 
	citecolor   = darkblue 
}
\usepackage{enumitem} 
\usepackage{titlesec} 
\titleformat{\section}
{\bfseries\centering\uppercase}
{\thesection.}
{1em}{}

\newcommand{\ee}{\end{equation}}
\newcommand{\be}{\begin{equation}}
\newcommand{\p}[1]{\left( #1 \right)}
\newcommand{\br}[1]{\left[#1\right]}
\newcommand{\cbr}[1]{\left\{#1\right\}}
\newcommand{\hc}[1]{#1^\dagger} 
\newcommand{\Hc}{\textrm{Hc}}
\renewcommand\bra[1]{{\langle{#1}|}}
\makeatletter
\renewcommand\ket[1]{%
	\@ifnextchar\bra{\k@t{#1}\!}{\k@t{#1}}%
}
\newcommand\k@t[1]{{|{#1}\rangle}}
\makeatother
\newcommand{\kb}[2]{\ket{#1} \bra{#2}}

\newcommand{\ra}{\rightarrow}
\newcommand{\om}{\omega}
\newcommand{\g}{\gamma}
\newcommand{\tw}{\tilde{t}}
\newcommand{\bin}{b_\textrm{in}}
\newcommand{\bout}{b_\textrm{out}}
\newcommand{\sm}{\sigma^-}
\newcommand{\spl}{\sigma^+}
\newcommand{\calT}{\mathcal{T}}
\newcommand{\Heff}{H_\textrm{eff}}
\newcommand{\Hsys}{H_\textrm{sys}}


\begin{document}
	
	\title{Quantum state transfer and input-output theory with time reversal} 
	\author{Kevin Randles}
	\email{krandles@uoregon.edu}
	\author{S.~J.~van Enk}
	\email{svanenk@uoregon.edu}
	\affiliation{Department of Physics and Center for Optical, Molecular and Quantum Science, University of Oregon, Eugene, OR 97403, USA}
	
	\begin{abstract}
		Being able to reliably transfer the quantum state from one system to another is crucial to developing quantum networks. A standard way to accomplish this transfer of information is by making use of an intermediate information carrier (e.g., a photon) that is emitted by the first system and absorbed by the second. For such a scenario one can develop an effective description by eliminating the intermediate degrees of freedom, which yields an effective direct coupling between the two systems. If, however, the spectral properties of the two systems are different, the photon's time-frequency shape 
		needs to be appropriately modified before it reaches the second system. We study here the effective description that results when we thus manipulate the intermediate photon.
		We examine a unitary transformation, $U$, that time reverses, frequency translates, and stretches the photon wave packet. We find that the concomitant modifications to the effective description can best be understood in terms of a change to the state's time argument, $\rho(t) = \rho_1(\tw) \otimes \rho_2(t)$, where $\tw$ is a fictitious time for the first system  that is stretched and runs backward. We apply this theory to three-level $\Lambda$-systems inside optical cavities, and we numerically illustrate how performing the unitary transformation $U$ results in improved quantum state transfer.
	\end{abstract}
	
	\maketitle	
	
	\newcounter{acounter}\refstepcounter{acounter}	
	\section{Introduction}
	\indent

	Quantum processors can be interconnected to make hybridized and distributed quantum devices that are more powerful than their isolated components \cite{kimble2008quantum,ritter2012elementary,hybrid21}.
	Hybridization allows for the individual system's strengths to be leveraged so as to create more versatile, robust, and scalable quantum information architectures \cite{hybrid06,hybrid15,hybrid15b,hybrid17,hybrid19,hybrid20,hybrid20b,hybrid21,cacciapuoti2019quantum,awschalom2021development}. For instance, one could use fast gate solid-state qubits as processors and long coherence time trapped-ion qubits for memory storage \cite{hybrid15}. 
	Distributed quantum information processing is essential to quantum networking \cite{cirac1997qst,ritter2012elementary,axline2018demand,kimble2008quantum}.
	A crucial part of this interconnection is achieving quantum state transfer (QST) between different quantum systems that are spatially separated \cite{transduction10,transduction20,transduction20b,maring2017photonic}. This transfer can be achieved using flying qubits, envisaged here as photons but phonons could be used for optomechanical systems \cite{hybrid20, transduction20b,stannigel2012optomechanical}, 
	to carry quantum information from one system to another \cite{cirac1997qst,kimble2008quantum,northup2014quantum}.
	Such direct transfer can be used to distribute entanglement, e.g., by sending one qubit of a Bell state to another location, with applications in quantum networking, quantum key distribution, and other protocols \cite{kimble2008quantum,cacciapuoti2019quantum}.
	
	Connecting different quantum systems with high fidelity is challenging because they may have different resonance frequencies and decay rates. In such cases it is well known that one ought to shift the frequency of the light pulse emitted by one system so as to interact resonantly with the receiving system and also stretch or compress the pulse in the time domain to match the receiving system's time scales \cite{qfc18,qfc21,maring2017photonic,shaping17}.
	 In addition, one should time-reverse the light pulse, simply because the light pulse that will be optimally absorbed by a system should be the time-reverse of what it would emit \cite{perfectexcitation09,timereversal14,timereversal}. Thus, this challenge can be dealt with by manipulating the light emitted by one system so that it will be absorbed by the next. 
	
	Analyzing the state transfer process requires a theoretical description of how light drives quantum systems. The description of how a coherent state interacts with an atom is straightforward and well known \cite{mollow1969power}. In contrast, describing how a general quantum state of the electromagnetic field interacts with an atom is surprisingly difficult. In 1993, Gardiner and Carmichael developed a strategy that allows one to manage a wider class of field states, called input-output theory \cite{gardiner1985input,gardiner1993driving,carmichael1993quantum}. In this theory, one includes a quantum description of the source of the light, and then assumes the light propagates freely and unidirectionally to a cascaded quantum system. 
	That photonic degree of freedom can then be eliminated (by solving the Heisenberg equations for the continuous mode bath operators) to obtain an \textit{effective description} where the source and receiver are directly coupled. The effective Hilbert space is then typically no longer infinite dimensional and in fact it is rather small if we can model the systems as ones with just a few energy levels (which we can if the interaction is nearly resonant).

	Employing this formalism, we consider two quantum systems, labeled `1' and `2,' that are coupled indirectly by the ability to exchange photons. We assume this coupling is unidirectional, that is, system 1 can produce a photon that travels to system 2, but no photon travels from system 2 back to system 1. 
	There is no reflection along the quantum channel linking the systems because we assume that the output field of system 2 is directed down a loss channel consisting of a different spatial mode; this could be implemented using a ring cavity (see Sec.~\ref{sec:example}) or circulator \cite{axline2018demand}.
	This unidirectional character of the theory is clear in the effective
	description, in which there are two types of time evolution:
	One is by discrete jumps and the other is continuous time evolution governed by a Schr\"odinger-like equation, but with a non-Hermitian effective Hamiltonian \cite{dalibard1992wave, gardiner1992wave, molmer93}. This effective Hamiltonian will contain a term proportional to $\sm_1 \otimes \spl_2$ without the Hermitian conjugate term (see App.~\hyperref[appA]{I}). Here $\sigma_j^\pm$ are the creation and annihilation operators for system $j=1,2$. 
	That is, there is a term that annihilates an excitation in system 1 and creates an excitation in system 2, but not vice versa.

	In this paper our primary motivation concerns what changes in the theoretical description if the light does not simply propagate freely but is manipulated by means of a unitary transformation $U$.
	This extends the seminal theoretical work of Cirac and co-workers \cite{cirac1997qst}, which describes a method for achieving ideal QST between identical atoms in cavities using time-symmetric pulses. The scheme that Ref.~\citenum{cirac1997qst} proposes, and likewise the scheme we develop, is reversible, so crucially the state can be transferred from an atom to the field and vice versa (field to atom). This reversible mapping has been experimentally verified \cite{boozer2007reversible}. 
	
	One question we answer is, for the class of unitary transformations, $U$, we consider, can we still eliminate the light from our description to  get a simple effective direct coupling between the two systems?
	The second question we answer is more subtle: The fact that the effective Hamiltonian contains just a term proportional to $\sm_1 \otimes \spl_2$ (naively) seems to indicate that the photon emitted by system 1 will be absorbed by system 2 even without manipulating the photon. However, we know that this alone does not suffice. In fact, the optimal sort of wave packet that would be absorbed by system 2 would have to be the time inverse of the wave packet system 2 would emit itself. So suppose we apply a unitary transformation that time reverses the wave packet emitted by system 1. Then, how can we see from the effective description that system 2 will absorb the photon more efficiently?
	
	We obtain the answers by slightly changing the standard interpretation of the equations obtained in the effective description 
	where the two systems are directly coupled.
	Doing that will lead us to define a mathematical object $\rho(t)$ consisting of two parts. One part is simply the state of system 2 at time $t$ and the other is the state of a fictitious system $\tilde{1}$ at a time $\tw = f(t)$ that decreases as a function of $t$ if we apply a time-reversal transformation. The time evolution of that mathematical object thus corresponds to system 2 evolving forward in time in a standard way, 
	yet it is driven by the backward-evolving fictitious system $\tilde{1}$.
	This is similar in spirit to the theory of the ``past quantum state'' \cite{gammelmark2013past}, which likewise introduces a mathematical object consisting of one forward-evolving standard quantum state and a backward-evolving part. In the latter case the backward-evolving part describes retrodiction, whereas in our case it results only if we time reverse the single-photon wave packet.  
	
	In Sec.~\ref{sec:setup} we rehearse the input-output methods of Gardiner \cite{gardiner1993driving} to model the two systems interacting unidirectionally via a bosonic bath connecting them. We determine the field resulting from the systems' interaction, which is the physical entity we transform in Sec.~\ref{subsec:fieldTrans}.
	In Sec.~\ref{sec:fieldTrans} we describe and analyze a scheme for QST using photon manipulation that could be used for hybrid quantum networking. Modeling the photon manipulation as a unitary transformation, we show how to include field transformations in input-output theory. We analyze how the field is transformed and how equations of motion (EOMs)  for system 2 operators are affected. 
	In Sec.~\ref{sec:example} we consider two systems each comprised of a three-level atom in a cavity, where a resonant two-photon Raman process is used to couple each atom to its respective cavity. Here we show that our scheme can be used to achieve ideal QST.
	We discuss our results and give some conclusions in Sec.~\ref{sec:discussion}. This includes a discussion of the  dynamics of the composite state of systems $\tilde{1}$ and 2, $\rho(t)$, where we add a new interpretation of the state's time argument.

	\section{General Setup}\label{sec:setup}
	We assume that each system ($j=1,2$) encodes the state of a qubit in a long-lived effective two-level system with ground state $\ket{g_j}$ and excited state $\ket{e_j}$. We suppose that system 1 starts in an arbitrary superposition $c_g \ket{g_1} + c_e \ket{e_1}$, which is the state we want to transfer, and that system 2 starts in its ground state $\ket{g_2}$. 
	Thus, we want to implement the transformation
	\be\label{eq:qubitTrans}
		\br{c_g \ket{g_1} + c_e \ket{e_1}} \otimes \ket{g_2} 
		\longmapsto 
		\ket{g_1} \otimes \br{c_g \ket{g_2} + c_e \ket{e_2}}. 
	\ee
	For now we assume that we have a way of inducing absorption and emission in these two-level systems via laser pulses. 
	To determine the requisite details of these pulses one needs to specify what the underlying systems encoding the qubits are, which we keep arbitrary for now.
	In Sec.~\ref{sec:example} we show, following the work of Ref.~\citenum{cirac1997qst}, how this can be done for three-level atoms in optical cavities, where the atomic ground states are coupled by a Raman transition, forming our qubits. 

	Based on the input-output formalism we model the quantum channel as a quasi-one-dimensional bath, with boson annihilation operators $b(\om)$, that both systems 1 and 2 are coupled to.
	Let the positions of systems 1 and 2 be $x_1 = 0$ and $x_2 = c \tau > 0$, respectively. Based on the formalism developed in Refs.~\citenum{gardiner1992wave} and \citenum{gardiner1993driving} we model the total Hamiltonian as
	\be\label{eq:Hamiltonian}
		H = \Hsys + H_\textrm{B} + H_\textrm{int},
	\ee
	where $\Hsys$ is the sum of the Hamiltonians for the two systems, and 
	\be\label{eq:HBath}
	H_\text{B} = \int d\om \abs{\om} \hc{b}(\om) b(\om)
	\ee
	and
	\begin{align}
		H_\text{int} &= i \int d\om 
		\Big\{
		\kappa_1(\om) \br{\sigma_1^- \hc{b}(\om) - \sigma_1^+ b(\om)} \nonumber\\
		&+ \kappa_2(\om) \br{\sigma_2^- \hc{b}(\om) e^{-i \om \tau} - \sigma_2^+ b(\om) e^{i \om \tau}} 
		\Big\}
	\end{align}
	are the Hamiltonians for the bath and its interaction with the two systems (in the rotating wave approximation), respectively ($\hbar =1$). 
	Integrals without explicit bounds are taken to be from $-\infty$ to $\infty$. 
	For much of our analysis $\Hsys$ is left unspecified and can be interpreted as a general operator on the two systems  (see Sec.~\ref{sec:example} for a concrete example).
	We work in the Heisenberg picture, so the operators $\sigma_j^\pm$ and $b$ depend on time. Note this is a quasi-one-dimensional setup in that the only propagating degree of freedom is the longitudinal mode characterized by $b(\om)$ and the other three quantum numbers (polarization and two transverse spatial modes) are fixed.
	
	We now make the Markov approximation of a flat coupling $\kappa_j = \sqrt{\g_j/2 \pi}$ within a narrow bandwidth of positive $\omega$ (e.g.,~the linewidth for optical cavities). This includes setting $\kappa_j(\omega)=0$ for negative $\omega$, encoding the unidirectional nature of the coupling \cite{gardiner1993driving,gardiner1992wave}. (In particular, see Sec.~II B of Ref.~\citenum{gardiner1992wave} for a justification of this approximation and some discussion of the integration limits.)
	Then we define the forward-traveling ‘photon field’ operator\footnote{We take this name for the field operator $A^+$ from Ref.~\citenum{timereversal}. Here $A^+$ is a vector potential scaled such that $A^+ \hc{\p{A^+}}$ is a photon flux and under the Markov approximation \cite{gardiner1992wave} it is proportional to the electric field (which is what Ref.~\citenum{gardiner1993driving} refers to it as).} as
	\be
	A^+(x,t) = \frac{1}{\sqrt{2 \pi}} \int d\omega b(\omega, t) e^{i \omega x/c}.
	\ee
	After solving the Heisenberg EOM for $b(\om ,t)$ with initial condition at $t_0 < t$, this field 
	can be expressed in terms of the rightward- and leftward-propagating `in fields,'
	\be 
	\bin^r(t) = \frac{1}{\sqrt{2 \pi}} \int_0^\infty d\om e^{-i \om (t-t_0)} b(\om, t_0)
	\ee
	and
	\be 
	\bin^l(t) = \frac{1}{\sqrt{2 \pi}} \int_0^\infty d\om e^{-i \om (t-t_0)} b(-\om, t_0),
	\ee
	respectively.\footnote{Note that $\bin^\textrm{total} = \bin^r + \bin^l \propto \int_{-\infty}^\infty d\om~e^{-i |\om| (t-t_0)} b(\om, t_0) $, where the $|\om|$ comes from Eq.~(\ref{eq:HBath}) \cite{gardiner1993driving}.}
	Namely,  one finds 
	\begin{align}\label{eq:inField}
		A^+(x,t) &= \bin^r\p{t_-} + \bin^l(t_+)
		+ \sqrt{\g_1} u(x) \sigma_1^-\p{t_-} \nonumber \\
		&+ \sqrt{\g_2} u(x - c\tau) \sigma_2^-\p{t_-+\tau},
	\end{align}
	where $t_\pm = t \pm x/c$ and $u(x)$ is the Heaviside step function with $u(0) = 1/2$. Using a ``final condition" at $t_1 > t$, one finds a similar expression for $A^+(x,t)$ in terms of the `out fields' $\bout^{r,l}$ defined similarly to $\bin^{r,l}$ with $t_0 \ra t_1$. Equating these expressions  for $A^+(x,t)$ 
	and using the independence of $x$ and $t$ then implies 
	\be\label{eq:leftPropFields}
	\bout^l(t) - \bin^l(t) = 0
	\ee
	and 
	\be\label{eq:rightPropFields}
	\bout^r(t) - \bin^r(t) = \sqrt{\g_1} \sigma_1^-(t) + \sqrt{\g_2} \sigma_2^-(t + \tau).
	\ee
	Physically, Eq.~(\ref{eq:leftPropFields}) says that the leftward-propagating free fields are not changed and Eq.~(\ref{eq:rightPropFields}) says that the rightward-propagating free fields are changed by the radiation from the two systems (as is consistent with the unidirectional coupling).

	As the leftward-propagating free field is unchanged, we focus on the rightward propagating field and abbreviate $b_\textrm{in,out} \equiv b_\textrm{in,out}^r$. 
	Then with 
	\be
	c_1(t) := \frac{\g_1}{2}\sm_1(t) + \sqrt{\g_1} \bin(t)
	\ee
	and 
	\be
	c_2(t) := \frac{\g_2}{2}\sm_2(t) + \sqrt{\g_1 \g_2} \sm_1(t - \tau) 
	+ \sqrt{\g_2} \bin(t - \tau),
	\ee
	the general Heisenberg EOM for a system operator $s(t)$ is \cite{gardiner1993driving}
	\be\label{eq:genEOM}
	\dot{s}(t) 
	= \sum_{j=1}^2 \p{
		\hc{c_j} \br{s,\sm_j} 
		- \br{s, \spl_j} c_j}
	-i \br{s, H_\text{sys}},
	\ee
	in which each operator has the same time argument $t$ and $[\cdot, \cdot]$ is the commutator. Note that $\bin^l$ does not appear in this expression due to the assumption that $\kappa(\om)=0$ for negative $\omega$, which effectively restricts the $\om$  integral bounds.
	
	Let $s_j$ denote a system $j$ operator. Then as $\br{s_1, \sigma_2^\pm} = 0$, Eq.~(\ref{eq:genEOM}) implies
	\be\label{eq:genEOM1}
	\dot{s}_1(t) = \hc{c}_1 \br{s_1,\sm_1}
	- \br{s_1, \spl_1} c_1
	-i \br{s_1, H_\text{sys}},
	\ee
	so $\dot{s}_1(t)$ depends on system 1 and field operators at the same time $t$. Similarly, as $\br{s_2, \sigma_1^\pm} = 0$, 
	\be\label{eq:genEOM2}
	\dot{s}_2(t)  = \hc{c}_2  \br{s_2, \sm_2}
	- \br{s_2, \spl_2} c_2
	-i \br{s_2, H_\text{sys}},
	\ee
	so $\dot{s}_2(t)$ depends on system 2 operators at the same time $t$ and on the delayed output of system 1 at time $t - \tau$ through $c_2(t)$. 
	
	As $s_1$'s EOM depends on a single time we can shift this time to $t-\tau$ to match the time arguments of the system 1 and field operators in $s_2$'s EOM. Using this structure, we can therefore write the composite state of both systems in the Schr\"odinger picture as 
	\be\label{eq:compState}
	\rho(t) :=\rho_1(f(t))\otimes \rho_2(t),
	\ee 
	with $f(t)=t-\tau$. 
	The point is that both the equations for system 1 and system 2 can refer to the same time $f(t)$ for the state of system 1. 
	Importantly, this is true both for Gardiner's original equations \cite{gardiner1993driving} and for our transformed equations (see App.~\hyperref[app:transInterpretation]{I\,B}), where $f(t)$ will take a more complicated form.	
	
	\section{Incorporating Photon Manipulation}\label{sec:fieldTrans}
	\subsection{Set up}
	As the systems are spatially separated, the QST of Eq.~(\ref{eq:qubitTrans}) happens indirectly via the exchange of a photon. The \emph{idea} is that laser pulses induce a coherent superposition of an emitted photon (if system 1 is excited) and the vacuum (if system 1 is in the ground state). This transfers the state to the field mode, where the coherent superposition then propagates along the channel to system 2.
	We then want laser pulses to transfer the state of the field to system 2, inducing absorption if there was a photon.
	 
\begin{figure}[h]
	\includegraphics[width=\linewidth]{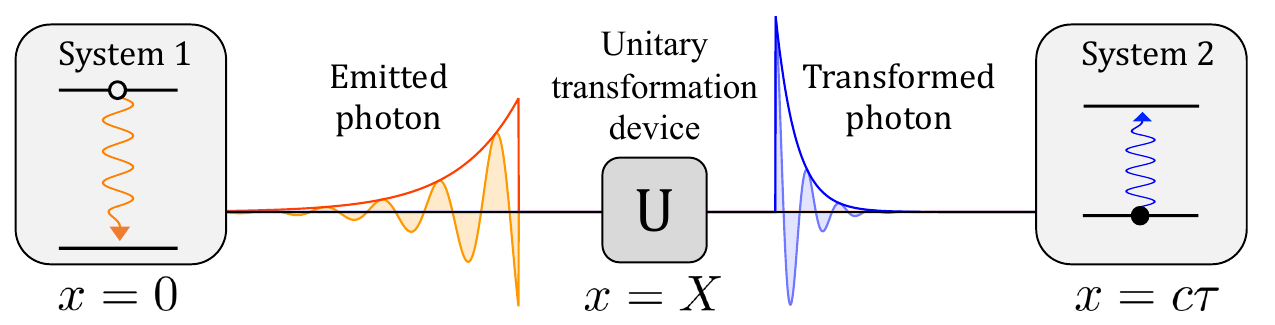}
	\caption{
		Scheme to transfer quantum information between distinct quantum systems along a channel. 
		For the sake of illustration, in this figure, we suppose that system 1 is initially excited and emits an exponentially decaying wave packet, which is common to spontaneous emission processes \cite{perfectexcitation09}.
	}
	\label{fig:schematicLabeled}	
	\end{figure}

	The \emph{problem} is that a photon emitted by system 1 will not necessarily interact strongly with and be absorbed by system 2. In fact, the optimal photon wave packet for system 2 to absorb is the time-reversed wave packet a lone system 2 would emit \cite{cirac1997qst}. 
	Thus, we consider applying a unitary transformation along the channel that time reverses, frequency translates, and stretches the photon wave packet, transforming it into the time-reversed output of system 2.
	With an appropriately implemented unitary, carefully designed laser pulses would induce absorption in system 2, thus completing QST, as illustrated in Fig.~\ref{fig:schematicLabeled}. 
	
	Transformations of this form could be physically implemented in a single 
	device, as proposed by Ref.~\citenum{timereversal}, or by applying different transformations sequentially, say, a time lens for time reversal and stretching \cite{shaping17,joshi2022picosecond}, followed by quantum frequency conversion. 
	The scheme proposed by Ref.~\citenum{timereversal} relies on nonlinear optical processes, either sum-frequency generation (SFG) or four-wave-mixing Bragg scattering, driven by a short classical pump pulse.
	
	Ref.~\citenum{cirac1997qst} developed a scheme to achieve QST between distant identical atoms by producing time-symmetric photon wave packets. Incorporating photon manipulation into our scheme makes it more versatile so that we can transfer information between systems with distinct spectral properties \cite{muller2017spectral} and the wave packets need not be time symmetric \cite{leuchs2012time}.
	[For completeness we note that Ref.~\citenum{timereversal14} demonstrates a method for producing a photon wave packet with a rising exponential shape, thus mimicking a time-reversed wave packet, by producing two photons and measuring one of them to herald the other. It is not clear whether such a heralding-based method could be adapted to our context, in which we need an appropriate superposition of the vacuum state and a single-photon state (conditioned on the state of system 1) in order to transmit the quantum state of one material system to another.] 
	
	To narrow our focus to the effect of the transformation, not its physical implementation, we will treat the transformation device as an idealized black box located at position $x=X$, $0< X < c \tau$.
	We discuss the correspondence between this abstracted device and a potential physical implementation using SFG \cite{timereversal} in App.~\hyperref[appB]{II}. 
	Thus, we compound the photon manipulations into a single unitary transformation 
	\be\label{eq:UTrans}
	U(\nu, \nu') = \frac{e^{i \nu T}}{ \sqrt{\xi}} \delta\p{ \nu' + \frac{\nu - \om_0}{\xi} } ,
	\ee
	which acts on a function in frequency space by time reversing it, scaling it by $\xi^{-1/2}$ (such that it will scale the photon flux by $\xi$ in the time domain), and shifting it by $\om_0$: 
	\be\label{eq:genfTrans}
		\tilde{F}(\nu) = \int d\nu' U(\nu, \nu') F(\nu') 
		= \frac{e^{i \nu T}}{\sqrt{\xi}} F\p{\frac{\om_0-\nu}{\xi}} .
	\ee
	It satisfies the unitarity condition
	\begin{align}
		\int d \nu'' U(\nu, \nu'') U^*(\nu', \nu'') 
		&= \delta(\nu - \nu').
	\end{align}
	We start in frequency space because the transformation would be physically  implemented using nonlinear-optical processes that modify frequencies \cite{timereversal,allgaier2017highly}. 
	
	The transformation can be expressed in the more intuitive time domain using Fourier transforms as 
	\be\label{eq:UTransTime}
	U(t, t') = \sqrt{\xi} e^{i \omega_0 (T-t)} \delta(t'-\tw_\xi )
	\ee 
	with
	\be
		\tw_\xi(t) := \xi(T-t),
	\ee
	where it acts on functions 
	as
	\be\label{eq:timeTrans}
		\tilde{F}(t) = \int dt' U(t, t') F(t').
	\ee
	The time parameter $T$ in the phase sets the time the transformation begins, as justified in Sec.~\ref{subsec:fieldTrans}. Thus, the parameters $\xi$, $\om_0$, and $T$ can be tuned such that the transformation time reverses the slowly varying envelope of the wave packet and shifts the resonance frequency and decay rate of system 1 to those of system 2.
	
	As the transformation of the photon wave packet occurs over a finite duration (whose start and end times have to be tuned),
	we may distinguish four different stages: 
	\begin{itemize}[leftmargin=2cm]
		\setlength\itemsep{0mm}
		\item [Stage 1] before transformation;
		\item [Stage 2] input field processing;
		\item [Stage 3] transformed field production;
		\item [Stage 4] transformation complete.
	\end{itemize}
	The initial field produced by system 1 freely propagates during Stage 1 until some time $t=t_i$, at which point the first part of the field to be transformed enters the transformation device at $x=X$. 	
	(We interpret these as stages of the wave packet transformation
	at the transformation device, though they can also be thought of as stages at system 2 with the appropriate time delay $\tau - X/c$.)
	Let $l$ denote the width of the input field we want to transform with corresponding duration $t_l = l/c$. 
	Then the transformation device will `process' this portion of the input field (which is ideally a single-photon wave packet) during Stage 2, $t_i < t < t_s  \equiv t_i + t_l$. 
	This results in a gap\footnote{The buffering of Stage 2 can be modeled as the activation of a $c$ mode in the vacuum state that mixes with our input field via a time-dependent beam-splitter transformation.	The $c$ mode has creation and annihilation operators $\hc{c_\textrm{in}}$ and $c_\textrm{in}$, respectively,  that satisfy the same commutation relations as $\bin$. 
		Then the field being transmitted through the unitary transformation device at $x=X$ is $\cos \theta(t) \bin(t) - i \sin \theta(t) c_\textrm{in}(t)$, where $\theta(t)$ is a switching function that determines whether the $c$ mode is active. Thus, $\theta$ should be a smoothed out (for continuity) square wave that is $\pi/2$ during Stage 2 of the transformation and zero elsewise.}
	in the field of duration $t_l$ during which a vacuum field $V(x,t)$, $\expval{V} = 0$, is produced because our system is `buffering' until $t=t_s$ when the transformed field production begins. (See App.~\hyperref[appB]{II} for a physical description of this buffering stage.)
	
	The transformed wave packet is produced during Stage 3 for $t_s < t < t_f \equiv t_s+t_l/\xi$ as the wave packet is stretched by $\xi$ in the time domain. 
	During this field production we block any incident field from passing through $X$ to ensure that there is always a single $f(t)$ describing the time argument of the field (making it a well-defined function), as later discussed in App.~\hyperref[app:transInterpretation]{I\,B}. This will result in some loss, yet the times $T$ and $t_l$ can be tuned to capture almost the entirety of the initial wave packet, making the loss arbitrarily small. We analyze this loss explicitly for our example in Sec.~\ref{subsec:exLosses}. 
	After the transformation is complete at $t = t_f$, i.e.,~during Stage 4, any of the original field (including noise) passing through $X$ will be unchanged and the field will freely propagate.
	
	\subsection{Field transformation}\label{subsec:fieldTrans}
	Consider the part of the field in Eq.~(\ref{eq:inField}) that describes the radiation propagating from system 1 to system 2, $x_1 < x < x_2$,
	\be\label{eq:sys1InField}
		A_i^+(x,t) = A_i^+(t_-) := \bin\p{t_-} + \sqrt{\g_1}  \sm_1\p{t_-}, 
	\ee	
	where we recall that $t_- = t - x/c$.
	The first term on the right-hand side is the input to system 1 and the second is the radiation emitted by system 1.
	The subscript $i$ denotes that this is the \textit{initial} field, which will be transformed at $x=X$, before it is incident on system 2, according to Eq.~(\ref{eq:timeTrans}) as
	\be
	\tilde{A}_{X}^+(t) = \sqrt{\xi} e^{i \om_0 (T-t)}  A_i^+\p{X, \tw_\xi}.
	\ee
	We can now determine the interval of time for which this transformed outgoing field is valid. For the transformation to be causal the field must depend on past times, $t \geq \tw_\xi$, such that $t\geq t_s \equiv T/(1+\xi^{-1})$, which defines the physical meaning of $T$.
	
	The transformed field being produced during Stage 3 will freely propagate for $x \geq X$ after the transformation as 
	\begin{align}\label{eq:transA}
	\tilde{A}^+(x,t) &= \tilde{A}_{X}^+\p{t - \frac{x-X}{c}} \nonumber \\
	&= \sqrt{\xi} e^{i \om_0 (T -t_- - X/c)} A_i^+\br{\calT(t_-)},
	\end{align}
	with $\calT(t_-) := \tw_\xi(t_-) - (1 + \xi) X/c$ as shorthand. 
	To produce a field with a decay rate matching system 2 we select
	\be \xi = \g_2/\g_1. \ee
	Then, comparing the original and transformed fields at system 2, $x = x_2 = c \tau$, we find that the system 1 operators transform as 
	\be\label{eq:simpA1Trans}
		\sqrt{\g_1} \sm_1(t - \tau) \ra \sqrt{\g_2}
		e^{i \om_0 (T - t + \tau - X/c)} \sm_1\br{\calT(t - \tau)}
	\ee
	and similarly for $\bin$, giving us the transformed, $\tau$-delayed, operators that impact the evolution of system 2 operators via the $c_2(t)$ terms in Eq.~(\ref{eq:genEOM2}). 
	
	Thus, the field propagating between the systems during any Stage is 
	\begin{align}\label{eq:genSys1Field} 
	A^+(x, t) &= 
	\begin{cases}
	V(x, t), &x \ge X, t_i < t - \frac{x - X}{c} < t_s \\ 
	\tilde{A}^+(x,t), & x \ge X, t_s < t - \frac{x - X}{c} < t_f \\ 
	A_{i}^+(x, t), & \textrm{elsewise} 
	\end{cases}.
	\end{align}
	The first case in Eq.~(\ref{eq:genSys1Field}) captures the processing of the portion of the wave packet to be transformed in Stage 2, during which a vacuum field $V$ is produced (see footnote 3). The second case captures the production and free propagation of the transformed field. The third case captures the free propagation of the initial field. 
	
	
	For $x \geq X$, this general expression for the field  only depends on $x$ and $t$ in terms of the combined variable $t_-$, so we can write it in the form
	\be
		A^+(x\geq X, t) = A^+(t_-) = \chi(t_-) A_i^+(f(t_-)).
	\ee
	Henceforth taking $X=0$ for simplicity ($X \neq 0$ adds trivial shifts that obfuscate the expression), we have
	\begin{align}\label{eq:transFieldPrefactor}
		\chi(t) &= \begin{cases}
		\sqrt{\xi} e^{i \om_0 (T - t)}, & t_s < t < t_f \\
		1, & \textrm{elsewise} 
		\end{cases}
	\end{align}
	and
	\be\label{eq:fTimeArg}
		f(t) = \begin{cases}
		\textrm{undefined}, & t_i < t < t_s \\
		\tw_\xi = \xi(T-t), & t_s < t < t_f \\
		t, & \textrm{elsewise}
		\end{cases},
	\ee
	where $A^+_i(f = \textrm{undefined})$ is interpreted to be the vacuum $V$ (see footnote 3).
	

	
	\section{Three-level atoms in cavities}\label{sec:example}
	\subsection{Hamiltonians}
\begin{figure}[h]
	\includegraphics[width=\linewidth]{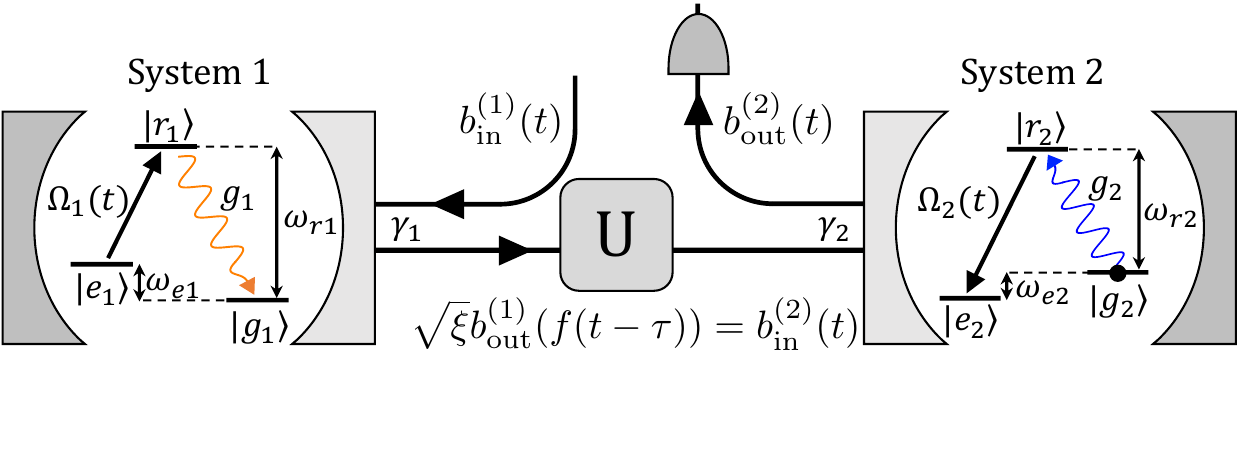}
	\caption{
		Schematic depiction of three-level atoms in cavities coupled via a quantum channel along which a unitary transformation is implemented (adapted from Ref.~\citenum{cirac1997qst}).
		The black arrows indicate laser pulses that induce emission in system 1 and absorption in system 2 (see the text for an explanation). We assume the input field to cavity 1 is in vacuum, $\big\langle b_\textrm{in}^{(1)} \big\rangle= 0$. 
	}
	\label{fig:exampleSchematic}	
	\end{figure}

	Now we will analyze a concrete example where each system, $j=1,2$, is comprised of a three-level atom, in the $\Lambda$-configuration with excited state $\ket{r_j}$ and ground states $\ket{g_j}$ and $\ket{e_j}$, that is strongly coupled to a mode of its respective high-$Q$ cavity with frequency $\om_{cj}$ and annihilation operator $a_j$. We take $\ket{g_j}$ to be the zero point of energy, $E_g = 0$, so 
	$\ket{r_j}$ has excitation frequency $\om_{rj}$ and the energy difference between $\ket{e_j}$ and $\ket{g_j}$ is $E_{ej} - E_{gj} = \om_{ej}$ (see Fig.~\ref{fig:exampleSchematic}).	
	This example and the underlying scheme extend and are based on the pioneering work of Ref.~\citenum{cirac1997qst} on QST in a quantum network. 
	What we change, relative to Ref.~\citenum{cirac1997qst}, is that the systems are not identical (they can have different parameters including resonance frequencies, $\om_{rj}$, and decays rates, $\g_j$) and we incorporate our unitary transformation into the scheme. 
	Additionally, Ref.~\citenum{cirac1997qst} assumed degenerate ground states as they had a Cs atom in mind. 
	As we want our scheme to apply more generally we do not assume the ground states to be degenerate ($\om_{ej} \neq 0$). This is a small change to the Hamiltonian but makes it applicable to a larger variety of implementations: Trapped ion qubits are generally not degenerate (e.g.,~two different hyperfine ground states or an \textit{S} ground state and a metastable \textit{D} state), superconducting qubits states are never degenerate, etc.
	
	For the benefit of the reader, we will relay the steps from Ref.~\citenum{cirac1997qst} that are crucial to our work. 
	The idea is that for each system, $\ket{e}$  and $\ket{g}$ are coupled by a Raman transition and form our qubits.
	Specifically, we drive an excitation from $\ket{e_1}$ to $\ket{r_1}$ using a laser pulse of frequency $\om_{L1}$ with Rabi frequency $\Omega_1(t)$ and phase $\phi_1(t)$. This is followed by the transition $\ket{r_1} \ra \ket{g_1}$ and the corresponding emission of a photon into cavity 1 with coupling $g_1$. 
	The photon leaks out of the cavity as a wave packet and propagates down the quantum channel and enters cavity 2. 
	Next, atom 2 undergoes the time inverse of the Raman process undergone by atom 1  (with different parameters $g_2$, $\om_{L2}$, $\Omega_2(t)$, and $\phi_2(t)$). The photon is then either
	absorbed, exciting atom 2 from $\ket{g_2}$  (where it is initialized) to $\ket{e_2}$, such that its state is transferred to atom 2 
	or 	 
	reflected down a different spatial mode (e.g.,~via a ring cavity geometry) and the transfer fails.	
	
	We want to design the first laser pulse (including the phase) to ensure that if there is an excitation in atom 1, it is transferred to cavity 1, and then to the channel. 
	Here we neglect any absorption in the transmission line, assuming that a photon emitted by system 1 will make it to system 2. 
	Though, analyzing deviations from this ideal case would be useful for understanding the scope of this work. 
	Furthermore, we do not want any other excitations to potentially be transferred to atom 2. Hence we assume a vacuum field input field to system 1, $\bin^{(1)} \ket{\textrm{in}} = 0$, where
	the input and output of each cavity are related by \cite{gardiner1985input,cirac1997qst}
	\be\label{eq:IOrelation} 
	b^{(j)}_\textrm{out}(t)	= b^{(j)}_\textrm{in}(t) + \sqrt{\g_j} a_j(t).
	\ee	
	Additionally, we want to ensure that the relative phase between $\ket{g}$ and $\ket{e}$ is transferred. This can be done using a local oscillator at the location of each system, $j=1,2$, with frequency $\omega_{ej}$ to serve as a well-defined frequency reference. We assume that the noise in each local oscillator is negligible and that the two local phases are stabilized relative to each other \cite{ball2016role}.
	Later, we adjust the Hamiltonians such that the two bare atomic states  effectively become degenerate, which makes the treatment more similar to Ref.~\citenum{cirac1997qst}.
	
	In \textit{our scheme}, the corresponding photon wave packet is manipulated while propagating between the systems, so the input to cavity 2 is the transformed  (and time-delayed) output of cavity 1.
	Thus, we also want to design the unitary transformation and the second laser pulse to minimize loss due to reflection at system 2, and hence achieve ideal QST. 
	The total system Hamiltonian is $\Hsys = H_1 + H_2$, where 
	\begin{align}
	H_j &= \om_{cj} \hc{a}_j a_j + \om_{rj} \kb{r_j}{r_j}  
	+ \om_{ej} \kb{e_j}{e_j} \nonumber \\
	&+ g_j \p{\kb{r_j}{g_j} a_j + \Hc} \nonumber \\
	&+ \frac{\Omega_j(t)}{2} \p{e^{i \br{\om_{Lj} t + \phi_j(t)}} 
	\kb{e_j}{r_j} + \Hc} 
	\end{align}
	is the Hamiltonian for system $j$ and Hc~denotes the Hermitian conjugate of the previous term.
	
	For each system, $j=1,2$, we go into rotating frames via transformations of the form $U = \exp\br{i \om_L t (\hc{a} a + \kb{r}{r}) + i \om_e t \kb{e}{e} }$, where the $\om_e \kb{e}{e}$ term is used to effectively make the two atomic ground states degenerate. 
	Then, assuming the lasers are strongly detuned $|\Delta_j| \gg \Omega_j, g_j, |\dot\phi_j|$ (with $\Delta_j = \om_{Lj} - \om_{rj}$) to suppress spontaneous emission, we can adiabatically eliminate the excited state $\ket{r}$ \cite{brion2007adiabatic}.
	Doing so, the new system Hamiltonians are 
	\begin{align}\label{eq:HjSimplified}
		H _j &= \bigg(\frac{g_j^2}{\Delta_j} \kb{g_j}{g_j} -
			\delta_j \bigg) \hc{a}_j a_j 
			+ \delta\om_j(t) \kb{E_j}{E_j}
		  \nonumber\\
		&\quad -i G_j(t) \p{a_j \kb{E_j}{g_j} - \Hc},
	\end{align}	
	where $\delta_j = \om_{Lj} - \om_{cj}$, $\delta \om_j(t) := \Omega^2_j(t)/4 \Delta_j$, and $G_j(t) := g_j \Omega_j(t)/2\Delta_j$ 
	are the Raman detunings, the ac Stark shifts to $\ket{E_j(t)} := e^{i \phi_j(t)} \ket{e'_j}$, and the Jaynes–Cummings interaction strengths for the effective two-level atoms, respectively \cite{cirac1997qst}.\footnote{We rephased $\phi_j \ra \phi_j + \pi/2$ so that the pulses $G_j$ are real. Then $\alpha_1$ and $\beta_1$ are real, while $\alpha_2$ and $\beta_2$ will be complex for $\zeta \neq 0$.} Here $\ket{e'_j} = U\ket{e_j} = e^{i \om_{ej} t } \ket{e_j}$ is the bare atomic state $\ket{e_j} $ in the new frame, meanwhile $\ket{g_j}$ is unchanged.
	
	\subsection{System evolution}
	Now we want to determine how the state of the systems, expanded as 
	\begin{align}\label{eq:psi}
		\ket{\psi(t)} = \alpha_1(t) \ket{Eg} \ket{00} + \alpha_2(t) \ket{gE} \ket{00} \nonumber \\
		+ \beta_1(t) \ket{gg} \ket{10} + \beta_2(t) \ket{gg} \ket{01},
	\end{align}
	will evolve. 
	The basis states, read from left to right, give the state of atoms 1, 2 ($E$ or $g$) and the states of cavities 1, 2 ($0$ or $1$).
	We leave out a $\ket{gg} \ket{00}$ term because, in the absence of noise, 
	the zero-excitation part of the transfer in Eq.~(\ref{eq:qubitTrans}),  $c_g \ket{g_1} \ket{g_2}\mapsto  c_g \ket{g_1} \ket{g_2}$, is trivial.
	Now we will go from the Heisenberg picture, in which system operators satisfy Eq.~(\ref{eq:genEOM}), to the Schr\"odinger picture. In doing so we will work with time-delayed operators and variables for system 1 (e.g.,~$\sm_1(t - \tau) \ra \sm_1(t)$  and so forth for $\bin$, $\Omega_1$, and $\phi_1$), effectively eliminating the time delay $\tau$.
	This Heisenberg to Schr\"odinger picture conversion can be described using the so-called quantum trajectory method, where $\ket{\psi}$ experiences smooth evolution governed by a non-Hermitian effective Hamiltonian $\Heff$ as well as discrete quantum jumps, where a so-called jump operator $J$ is randomly applied to the state of our two systems (then the state has to be renormalized $\ket{\psi} \ra J\ket{\psi}/\bra{\psi}\hc{J} J\ket{\psi}$) \cite{gardiner1992wave}. 
	Employing this method, we find that for a vacuum field input to system 1 the effective Hamiltonian in the rotating frame used above is
	\begin{align}\label{eq:exHeff}
	\Heff = H_1 + H_2 
	-&\frac{i}{2} \Big(\g_1 \hc{a}_1 a_1  + \g_2 \hc{a}_2 a_2  \nonumber \\ 
	&+ 2 \sqrt{\g_1 \g_2} e^{i \zeta t}\hc{a}_2 a_1 \Big),
	\end{align}
	where $H_{1,2}$ are the Raman Hamiltonians of Eq.~(\ref{eq:HjSimplified}) and $\zeta \equiv \om_{L2} - \om_{L1}$ is a frequency mismatch that degrades the QST when the unitary transformation is not applied. The corresponding phase factor, $e^{i \zeta t}$, comes from two different rotating frames for the two systems.
	(See App.~\hyperref[appA]{I\,A} for further discussion of the quantum trajectory method including a more general derivation of $\Heff$.)
	
	We find the state amplitude EOMs 
	\begin{subequations}
		\begin{align} 
		\dot\alpha_1 & = -G_1 \beta_1, \label{eq:EOMa}\\
		\dot\beta_1 & = G_1 \alpha_1 - \frac{\gamma_1}{2} \beta_1, \phantomsection\label{eq:EOMb} \\
		\dot\alpha_2 & = -G_2 \beta_2, \phantomsection\label{eq:EOMc} \\ 
		\dot\beta_2 & = G_2 \alpha_2 - \frac{\gamma_2}{2} \beta_2 - \sqrt{\gamma_1 \gamma_2} e^{i \zeta t} \beta_1, \label{eq:beta2EOMd}
		\end{align}
	\end{subequations}
	where we picked laser phases and detunings that satisfy $-\dot\phi_j \equiv \delta\omega_j$ (so $\phi_j$ is determined by the pulse $\Omega_j$ up to an initial condition)
	and $\delta_j = g_j^2/\Delta_j$ (so that we must pick specific laser frequencies $\om_{Lj}$\footnote{The condition $\delta = g^2/\Delta$ (for each system, $j=1,2$) implies that the laser detunings and hence laser frequencies are determined by
		$
		\Delta = \om_L - \om_r = \frac{\om_c - \om_r}{2} \pm \sqrt{\p{\frac{\om_c - \om_r}{2} }^2 + g^2}, 
		$
		where $|\Delta| \gg g$ so the $+$ solution is valid and implies
		$
		\om_L  = \om_c + \frac{g^2}{\om_c - \om_r} + \order{\frac{g^4}{(\om_c - \om_r)^3}}
		$
		such that we need to lase near the cavity resonance, $\om_L  \approx \om_c$.}), respectively. 
	We assume these conditions are met, that is, we assume Raman resonance once all shifts are taken into account, and do not consider  errors due to a nonzero Raman detuning. 
	Note that the EOMs for system 2 are the same as those for system 1 except for the $\beta_1$ term in Eq.~(\ref{eq:beta2EOMd}) which encapsulates how an excitation in cavity 1 is transferred to cavity 2. 
	
	\subsection{Photon transformation}
	So far, this section is just a generalization of Ref.~\citenum{cirac1997qst} to non-identical, non-degenerate systems.
	Our larger modification is incorporating the effect of the unitary transformation so that the two non-identical systems naturally interact with each other. Thus we will now analyze the impact of the unitary transformation on the amplitude EOMs.
	The photon wave packet emitted by system 1 can be specified by the mode function
	\be
		\Psi_1(t):=	\bra{gg}\bra{00} b^{(1)}_\textrm{out} \ket{\psi(t)}, 
	\ee
	whose modulus squared is the expectation value of the number operator for the output of cavity 1, $N_1 = \big(b^{(1)}_\textrm{out}\big)^\dagger b^{(1)}_\textrm{out}$, with respect to the state $\ket{\psi}$:
	\be
		\langle N_1(t) \rangle = |	\Psi_1(t)|^2.
	\ee
	For a vacuum field input 
	\be
			\Psi_1(t) =	\bra{gg}\bra{00} \sqrt{\g_1} a_1 \ket{\psi} 
			 =	\sqrt{\g_1} \beta_1(t).
	\ee
	
	The transformed (tilded) cavity creation operator can be found using Eq.~(\ref{eq:simpA1Trans}) at system 1 (again with $X = 0$) with $\sm_1 \ra a_1$ to be
		\be \tilde{a}_1(t) = \chi(t) a_1(f(t)). \ee
	It follows that the transformed wave packet is
	\be
		\tilde{\Psi}_1(t) = \bra{gg}\bra{00} \sqrt{\g_1} \tilde{a}_1 \ket{\psi} 
		 = \sqrt{\g_1} \chi(t) \beta_1(f(t))
	\ee
	so that system 2 is driven by the transformed amplitude
 	\be\label{eq:b1Trans}
 		\beta_1(t) \ra \tilde{\beta}_1(t) = \chi(t) \beta_1(f(t)),
 	\ee
 	whose form varies depending on what Stage of the transformation we are considering and we interpret $\beta_1(f = \textrm{undefined}) \equiv 0$.
	Hence under the transformation, Eq.~(\ref{eq:beta2EOMd}) becomes
	\be\label{eq:TransEOMd}
		\dot\beta_2 = G_2 \alpha_2 - \frac{\gamma_2}{2} \beta_2 - \sqrt{\gamma_1 \gamma_2} e^{i \zeta t} \tilde{\beta}_1
	\ee
	with the other equations (\ref{eq:EOMa}-\hyperref[eq:EOMc]{c}) remaining the same. 
	For $ t_s < t < t_f$ this gives
	\be
		\dot\beta_2 = G_2 \alpha_2 - \frac{\gamma_2}{2} \beta_2 - \gamma_2 e^{i \om_0 T} e^{i (\zeta - \om_0) t} \beta_1(\tw_\xi),
	\ee
	which motivates the choice $\om_0 = \zeta$ for the frequency shift of the transformation in order to cancel out the time-dependent phase in the last term of Eq.~(\ref{eq:TransEOMd}) during Stage 3.

	Now the problem is to find the pulse shapes $G_j \propto \Omega_j$ such that ideal QST occurs when we implement our unitary transformation, that is, we want
	\be\label{eq:coefficientBCs}
		 \alpha_1(-\infty) = |\alpha_2(+\infty)| = 1
	 \ee
	(with the other amplitudes equal to zero at these times due to normalization).   
	Here we are guided by the expectation that the laser pulse inducing absorption in system 2, $G_2$, must be time reversed and stretched relative to the pulse inducing the emission of system 1, $G_1$. Specifically, we let
	\be\label{eq:G2}
		G_2(t) = \xi G_1(\tw_\xi)
	\ee
	such that given solutions $\alpha_1$ and $\beta_1$ there are corresponding solutions $\beta_2(t) = - e^{i \om_0 T} \beta_1(\tw_\xi)$ and $\alpha_2(t) = e^{i \om_0 T} \alpha_1(\tw_\xi)$. 
	Note that we are not optimizing $G_2$ for every parameter value (starting time $t_i$, duration $t_l$, etc.). Rather, we select this pulse as it is agnostic to such details, working generally as long as $l$ is large enough for (almost) all of wave packet to be transformed.
	With this choice for $G_2$, constructing $G_1$ such that $\alpha_1(-\infty) = 1$ ensures that $|\alpha_2(+\infty)| = 1$ for sufficiently large $l$ (assuming both pulses are implemented correctly).
 
	To guarantee that the state's phase is transferred one must do two things: Use the local oscillators mentioned above to define clocks relative to which one can account for free precession, and account for the extra phase $e^{i \om_0 T}$ that appears in $\alpha_2$ due to the unitary.\footnote{We get the phase factor $\exp(i\omega_0 T)$ only when we apply the unitary (Stage 3). So if some of the untransformed portion of the photon is absorbed by system 2, we would induce a phase error. We assume either that the frequency shift $\omega_0 = \zeta$ is so large that the photon will not be absorbed without our unitary, or that we control $\omega_0$ and $T$ such that $\omega_0 T = 0$ (modulo $2\pi$) when $\omega_0$ is small.}
	This argument implicitly assumes that
	$\tilde{\beta}_1(t) = \sqrt{\xi} e^{i \om_0 (T - t)} \beta_1(\tw_\xi) \forall t$, which is only valid during Stage 3. However, if we transform enough of the wave packet (make $l$ big) $\beta_1$ will be negligibly small outside of Stage 3, making this an increasingly good approximation as $l$ increases.

	Now our task is to find a pulse $G_1$ such that the boundary conditions of Eq.~(\ref{eq:coefficientBCs}) are satisfied. From Eqs.~(\ref{eq:EOMa}--\hyperref[eq:EOMb]{b}) we find that $\alpha_1$ and $G_1$ uniquely determine each other. Thus, we can reframe the task as specifying an $\alpha_1$ that goes from 1 at $t = -\infty$ to 0  at $t = +\infty$,\footnote{Note that only some amplitude functions $\alpha_1(t)$ have a corresponding pulse $G_1(t)$ that produces them because the argument of the square root in Eq.~(\ref{eq:G1inTermsOfAlpha1}) must be positive as $G_1$ is real (the corresponding phase $\phi_1(t)$ has already been taken out). However, any positive $\alpha_1$ that monotonically decreases, as we have here, can be produced.} from which the first pulse is uniquely specified (up to a sign):
	\be\label{eq:G1inTermsOfAlpha1}
		G_1(t) =  \sqrt{\frac{\dot\alpha_1^2 e^{ \gamma_1 t}}{e^{ \gamma_1 t_p} \beta _1^2(t_p) - 2\int_{t_p}^t dt' e^{\gamma_1 t'} \dot\alpha_1(t') \alpha_1(t')}}.
	\ee
	Here $t_p$ is some early time by which the first atomic state is \emph{prepared} in the desired superposition state, $c_g \ket{g_1} + c_e \ket{e_1}$, so $\beta _1(t_p)$ should be zero.
	
	A natural choice of $\alpha_1$ is the logistic function
	\be\label{eq:logalpha1}
		\alpha_1(t) = \frac{1 + \tanh(- k t)}{2},
	\ee
	from which we can find the corresponding pulse $G_1$ via Eq.~(\ref{eq:G1inTermsOfAlpha1}) and $ \beta_1 = -\dot\alpha_1/G_1$ from Eq.~(\ref{eq:EOMa}). We assume the laser pulse applied to system 1, $G_1$, is implemented perfectly. 
	The ability to control the shape and timing of single-photon pulses has been demonstrated experimentally for similar systems \cite{keller2004continuous}.
	Then we can solve for $\alpha_2$ and $\beta_2$ numerically using Eqs.~(\ref{eq:EOMc}), (\ref{eq:TransEOMd}), and (\ref{eq:G2}). We plot the solutions and the corresponding laser pulses below in Fig.~\ref{fig:coefficentsInsetG}. 
	Additionally, we plot the photon wave packet's square modulus for the various stages of the transformation in Fig.~\ref{fig:UTransStages}. 
	In our plots we work in natural units based on the rate $\g_2$ and the speed of light $c$; so times are in units of $\g_2^{-1}$, lengths are in units of $c\g_2^{-1}$, etc.  

	\begin{figure}[h]
	\centering
	\includegraphics[width=\linewidth]{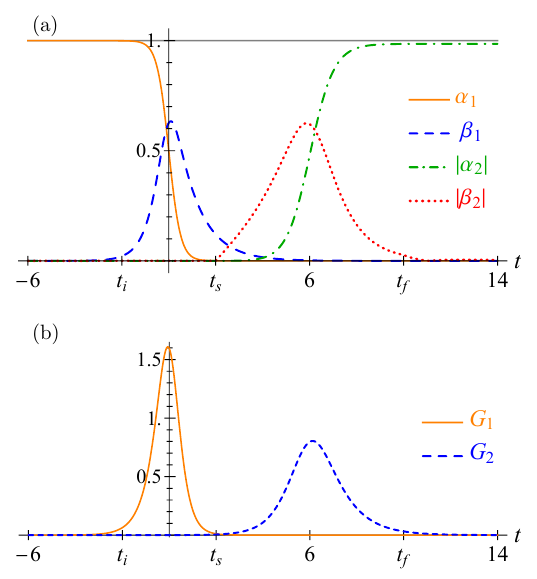}
	\caption{
		(a) Probability amplitudes and (b) corresponding laser pulses $G_{1,2}$  plotted as functions of time $t$ (all quantities are in units where $\g_2= 1$)
		for a logistic $\alpha_1$ with $k=2$ in Eq.~(\ref{eq:logalpha1}), taking $T=6$, $\xi = 1/2$, $\om_0 = \zeta = 50$, and $t_l=4$.
		Here atom 1 is initially excited, $\alpha_1^2(-\infty) = 1$, and this excitation is transferred to atom 2 with probability $|\alpha_2(\infty)|^2 = 0.97$ 
		($|\alpha_2|$ falls just below 1, the upper line, for large $t$). 
	}
	\label{fig:coefficentsInsetG}	
	\end{figure}

	\begin{figure}[h] 
	\centering 
	\includegraphics[width=\linewidth]{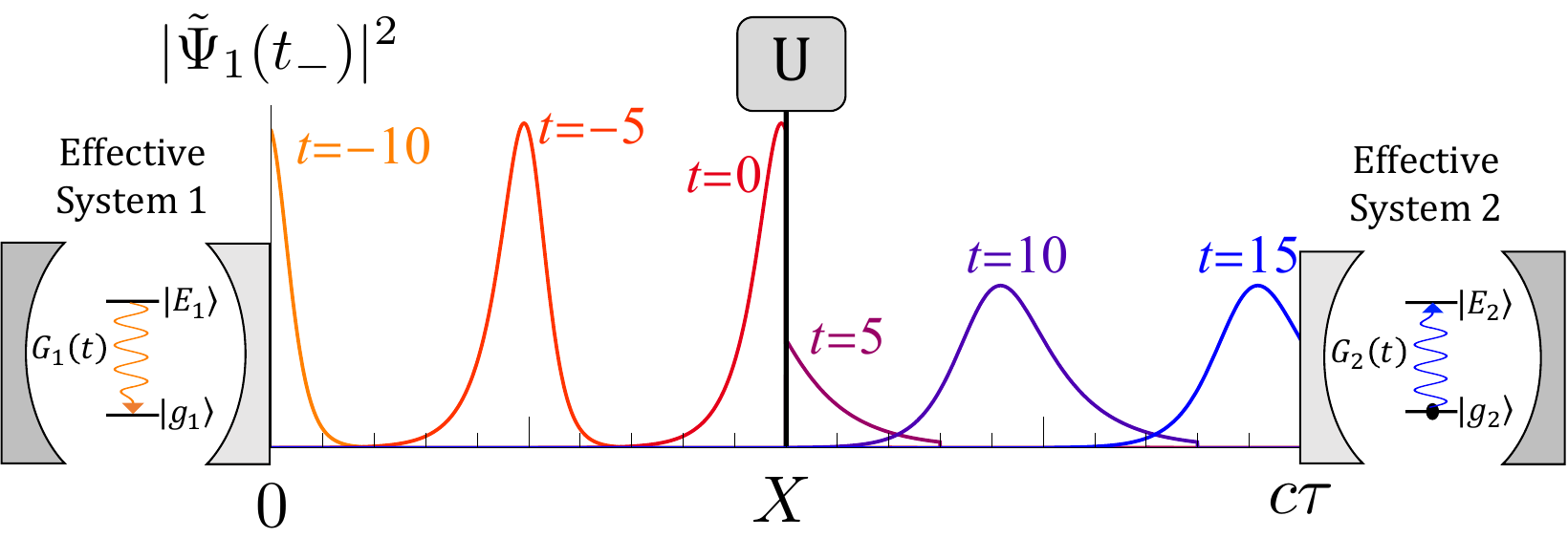}
	\caption{
		Plot of $|\tilde\Psi_1(t_-)|^2$ vs $x$ for various $t$ (in units of $\g_2^{-1}$), illustrating the propagation of the photon for a logistic $\alpha_1$ with the same parameters as in Fig.~\ref{fig:coefficentsInsetG}. 
	The wave packet propagates until $t_i = -2$, at which point it begins being processed until $t_s = 2$. The transformed wave packet is produced until $t_f = 10$, and then it propagates freely and induces absorption in system 2 with $97\%$ success rate. Here we let $X=c\tau/2$ for illustration.}
	\label{fig:UTransStages}
	\end{figure}
	
	\subsection{Two types of loss}\label{subsec:exLosses}
\begin{figure}[h] 
		\centering 
		\includegraphics[width=\linewidth]{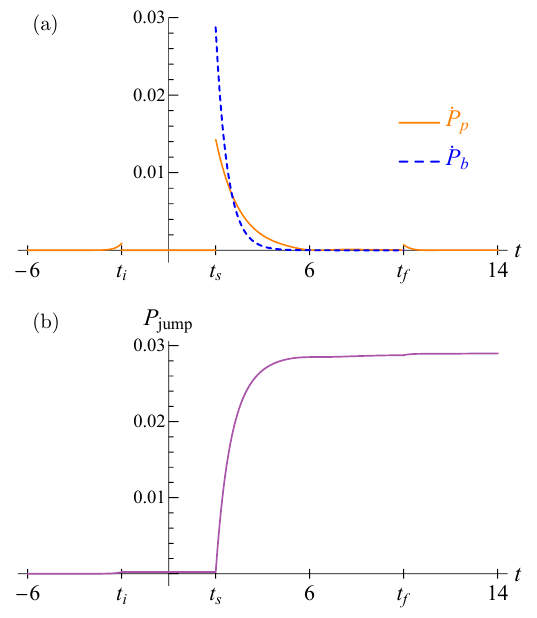}
		\caption{(a) Probability density functions for loss due to blocking (dashed line), which is only plotted during Stage 3 where it occurs, and due to imperfections in the shape of the photon wave packet (solid line).  
		(b) Probability of a quantum jump up to time $t$ (in units of $\g_2^{-1}$).
		The parameter values are the same as in Fig.~\ref{fig:coefficentsInsetG}. Note that $P_\textrm{jump}$ tends to $0.029$
		, as is consistent with the $0.97$ success rate we found previously.}
		\label{fig:PjumpVst}
	\end{figure}

	We can quantify whether we succeeded in transferring the quantum state from system 1 to 2 by analyzing two kinds of loss that are present in our scheme: 
	loss due to blocking the incident field from propagating through $x=X$ during Stage 3 
	and loss due to an imperfect photon wave-packet shape leading to the field being reflected from system 2 out another spatial mode, which occur with probabilities of $P_b$ and $P_p$, respectively. The probability density functions for these losses are
	\be
		\dot P_b= \begin{cases}
			 \g_1|\beta_1|^2, & t_s < t < t_f \\
			 0, & \textrm{elsewise}
		\end{cases},
	\ee 
	as we block the incident field during Stage 3, and 
	\be\label{eq:PpPdf}
			\dot P_p = \big| \sqrt{\g_1} e^{i \zeta t} \tilde{\beta}_1 + \sqrt{\g_2} \beta_2 \big|^2,
	\ee
	which comes from looking at the decay of the norm of the effective state due to the unitary transformation and accounts for the interference between the cavity amplitudes. Without the unitary, $\dot P_p$ is simply minus the time derivative of the norm state of $\ket{\psi}$.\footnote{One can equivalently write the probability density of loss due to an incorrect photon wave packet  as
		$
		\dot P_p = - \frac{d}{dt} \p{ \frac{|\tilde{\alpha}_1|^2 + |\tilde{\beta}_1|^2}{df/dt}
			+ |\alpha_2|^2 + |\beta_2|^2},
		$
		where the quantity in parentheses acts as a norm for system 2 and the fictitious system $\tilde{1}$ driving it. Here $\tilde{\alpha}_1$ is defined in a  manner analogous to $\tilde{\beta}_1$ in Eq.~(\ref{eq:b1Trans}).}
	 The total quantum jump rate up to time $t$ is then 
	\be
		P_\textrm{jump}(t) 
		= \int_{t_p}^{t} dt' \br{\dot P_b(t') + \dot P_p(t')}
	\ee
	as illustrated in Fig.~\ref{fig:PjumpVst}.

\begin{figure}[h] 
	 \centering 
	 \includegraphics[width=\linewidth]{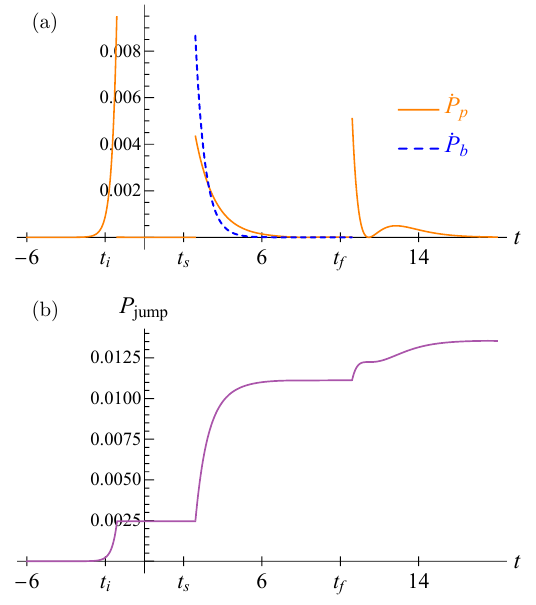}
	 \caption{Plots of (a) the probability density functions and (b) the total jump probability, choosing the ideal value of $T$ (keeping $\xi = 1/2$, $\om_0 = \zeta = 50$, and $t_l =4$) such that $t_i = -1.4$	(all quantities are in units where $\g_2= 1$).
	 	Here $P_\textrm{jump} \ra 0.014$ 
	 	for large $t$, which is slightly less than half of the loss probability we found previously with $t_i = -2$.
	 }
	 \label{fig:tiIdealJump}
	 \end{figure}

	For Figs.~\ref{fig:coefficentsInsetG}-\ref{fig:PjumpVst}  we selected the parameter values $T=6$, $\xi = 1/2$, $\om_0 = \zeta = 50$, and $l =4$ such that $t_i = -t_s = -2$ because $G_1$ is large during this window (see Fig.~\ref{fig:coefficentsInsetG}b). 
	This is a good choice but not the ideal choice, which for the same values of $\xi$, $\om_0$, $\zeta$, and $l$ involves choosing $T= 7.8$ such that $t_i = -1.4$. The corresponding results are plotted in Fig.~\ref{fig:tiIdealJump}, in which the errors are spread out among the stages due to delaying the start of the transformation from $t_i = -2 \ra -1.4$.
	 This change in starting time was beneficial as it led to larger portions of $\beta_1$ being transformed and hence more destructive interference between $\tilde{\beta}_1$ and $\beta_2$ in Stage 3 (see Eq.~(\ref{eq:PpPdf})).
	  
\begin{figure}[h] 
	\centering 
	\includegraphics[width=\linewidth]{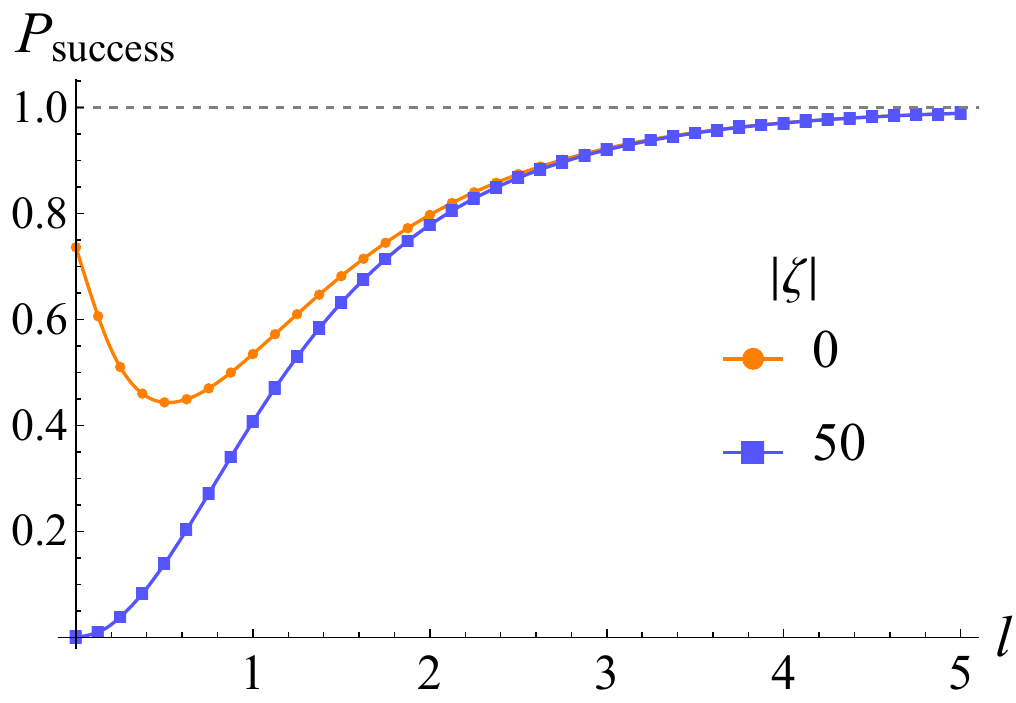}
	\caption{Plot of $P_\textrm{success}$ vs $l$ for identical laser frequencies, $\zeta=0$, and for a frequency mismatch of $|\zeta|=50$ (all quantities are in units where $\g_2= c = 1$). Here we use the ideal frequency shift $\om_0 = \zeta$, and $\xi = 1/2$, $T=3 t_l/2$ such that $t_s = t_l/2$. (See the text for further discussion.)
		} 
	\label{fig:PsuccessVsl}
	\end{figure}

	Both of these losses can be reduced by transforming more of the pulse. This can be accomplished by either increasing $l$ for a given $G_1$ or picking $G_1$ (or equivalently $\alpha_1$) such that the wave packet is more localized. 
	In Fig.~\ref{fig:PsuccessVsl} we illustrate how the probability of success, $P_\textrm{success} \equiv |\alpha_2(t \ra \infty)|^2$, 
	 approaches 1 as $l$ is increased. 
	 Note that if the first pulse is implemented correctly so that $\alpha_1 \ra 0$ for large times, then one finds that
	 $P_\textrm{success} = 1 - P_\textrm{jump}(t \ra \infty) = (|\alpha_1|^2 + |\beta_1|^2 + |\alpha_2|^2 + |\beta_2|^2)|_{t \ra \infty}$.
	 For $\zeta = 0$, the systems are already resonant so no frequency shift is needed and hence $P_\textrm{success}$ 
	decreases if only a small amount of the photon wave packet is transformed (as then much of it is blocked).

	We see that when the  frequencies of the lasers driving the systems differ appreciably, i.e.,~when $|\zeta|$ is large, $P_\textrm{success}$ is nearly 0 without the unitary (when $l =0$) and so performing a transformation is critical for QST.  
	Recall that, in our numerics, frequencies are measured relative to the cavity decay rate $\gamma_2$, which is typically on the order of kHz. Thus, `large $|\zeta|$' is relative to $\gamma_2$. Note that the laser frequency mismatch $\zeta$ will be similar in magnitude to the cavity resonance frequencies (see footnote 5) for which typical values are on the order of THz $=10^9$ kHz (this is consistent with our high-quality factor 
	assumption for each cavity).
	So generically $|\zeta| \gg \gamma_2 = 1$ (in our units) and then $P_\textrm{success}$ monotonically increases from 0 at $l=0$ to 1 for large $l$ (see the $|\zeta| = 50$ curve in  Fig.~\ref{fig:PsuccessVsl}, which does not change significantly as $|\zeta|$ is further increased;\footnote{For completeness, we note that there is $\zeta$-dependent interference  between  the cavity fields, $\exp(i\zeta t)\tilde{\beta}_1$ and $\beta_2$ (see Eq.~(\ref{eq:PpPdf})), which results in $P_\textrm{success}$ oscillating about the large-$|\zeta|$ limit. However, this effect is only pronounced for small but non-zero $|\zeta|$. For instance, for $|\zeta|=10$ the oscillations are visually discernable, while for $|\zeta| = 25$ they are not.} we keep $|\zeta|$ relatively small in our numerics to avoid issues with highly oscillatory integrals). 
	
	Experimentally, increasing $l$ would necessitate using more nonlinear optical media to control the unitary transformation, whereas tailoring $G_1$ to produce a narrow wave packet demands more precise control of the timing so that deviations from the ideal value of $T$ are more costly. 
	An analysis of the benefits and limitations of each of these solution methods is beyond the scope of the present work, as is consideration of loss due to a frequency mismatch $\om_0 \neq \zeta$. 
	
	\section{Discussion}\label{sec:discussion}
	In this work we have demonstrated that photon manipulation can be incorporated into input-output theory to achieve, in principle, ideal quantum state transfer (QST) between non-identical systems. This offers a potentially versatile scheme for QST in hybrid quantum networks. 
	We showed, using analytical and numerical means, how this scheme can be applied to systems of three-level atoms in cavities to achieve ideal QST. Our scheme could be generalized to systems with more levels, where more laser pulses and their time-reversed counterparts need to be used to induce emission and absorption in the systems, respectively \cite{gorshkov2007universal,giannelli2018optimal,biswas2021detecting}.
	Furthermore, our scheme could readily be adapted to other material systems, with the details contingent on the forms of the system Hamiltonians. 
	
\begin{figure}[h]
	\includegraphics[width=0.8\linewidth]{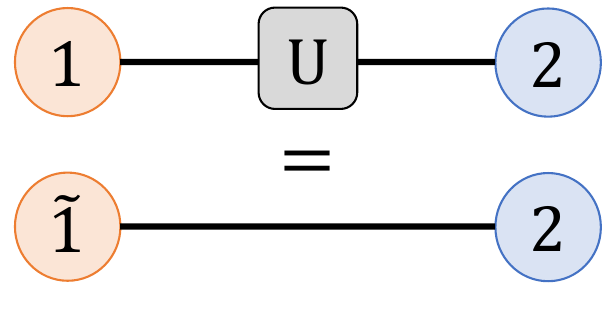}
	\caption{Schematic illustrating the interpretation of Eq.~(\ref{eq:transState}) (see the main text for details).
	}
	\label{fig:interpretationSchematic}
	\end{figure}

	A main result is a new interpretation of system 1's time argument in the effective description where the systems are directly coupled. In particular, we find that the mathematical object
	\be\label{eq:transState}
		\boxed{ \rho(t) = \rho_1(\tw)\otimes \rho_2(t) }
	\ee
	describes the composite state of system 2 coupled to a fictitious system $\tilde{1}$, which would produce the transformed wave packet. 
	Then we can eliminate the bath and unitary transformation from the description and obtain an evolution equation for the state $\rho$, in the standard form
	\be\label{eq:LindbladMasterEq}
	\dot{\rho} =  i \comm{\rho}{H_0} + \mathcal{L}[J]\rho 
	\ee 
	with 
	\be 
	\mathcal{L}[J]\rho  = J \rho \hc{J} 
	- \frac{1}{2} \cbr{\rho, \hc{J} J}
	\ee
	a Lindblad superoperator, $J$ a jump operator, and $\cbr{\cdot, \cdot}$ the anticommutator. 
	Yet crucially, the effects of the transformation (the scaling by a factor $\xi$ and time reversal) are accounted for by the change in the time argument of the state of system 1, $t \ra \tw = \xi (T-t+\tau)$, during the transformation (see App.~\hyperref[appA]{I} for more details).
	This has the interpretation that the unitary transformation can be incorporated into the effective description of the dynamics of system 2 by simply letting system 1 (which drives system 2) run backward in time at a new decay rate (see Fig.~\ref{fig:interpretationSchematic}).
	Thus, the effective description in the quantum trajectory method is unchanged, yet the interpretation of the state does change.
	
	Clearly, implementing the unitary transformation device in our scheme would be challenging. We discuss a nonlinear optical based method to implement our transformation in the optical domain in App.~\hyperref[appB]{II}. However, if we wanted to perform microwave to optical transduction we would need a different implementation of our transformation, e.g.,~using atomic ensembles or electro-optical systems for frequency conversion \cite{transduction20}, followed by our shaping and time-reversal procedure.
	Accordingly, modeling a more realistic, imperfect transformation device is an important extension of this work. We leave for future work the analysis of additional losses and limitations in the transformation.

	\section*{Appendix I: Schr\"odinger evolution}\phantomsection\label{appA} 
	\setcounter{equation}{0}
	\renewcommand{\theequation}{I\arabic{equation}}
	\subsection{Without transformation}
	The identity 
	\be
	\Tr(\rho_H\dot{s}(t))=\Tr(\dot{\rho}_S(t) s)
	\ee
	connects the Heisenberg picture operator and state on the left to the Schr\"odinger picture operator and state on the right. Using this identity, the cyclic nature of the trace, and Eq.~(\ref{eq:genEOM}), we find the evolution equation for the systems and field to be
	\be\label{eq:SPicEOM}
	\dot{\rho}_S = \sum_{j=1}^2 \p{
		\comm{\sm_j}{\rho_S \hc{c_j}} 
		- \comm{\spl_j}{c_j \rho_S}}
	-i \comm{H_\text{sys} }{\rho_S} .
	\ee
	In the Schr\"odinger picture the operators no longer evolve with time as any implicit time dependence is shifted onto $\rho_S(t)$.
	
	The transmission 
	can be described in an effective description where the two systems are directly coupled. This effective description is obtained by tracing over the bath, yielding the state of the systems 
	\be 
	\rho(t) = \Tr_\textrm{B} (\rho_S),
	\ee
	which is the same state as in Eq.~(\ref{eq:compState}). Considering coherent input states to the systems $\ket{\beta} \bra{\beta}$ such that $\bin \ket{\beta} = \beta \ket{\beta}$ with $\beta \in \mathbb{C}$ and
	\be
	\Tr_\textrm{B}(\bin \rho_S)  
	= \beta \rho,
	\ee
	we find that $\rho$ satisfies Eq.~(\ref{eq:SPicEOM}) with $\bin \ra \beta$. 
	After some algebra, we can write this master equation for $\dot{\rho}$ in the Lindblad form of Eq.~(\ref{eq:LindbladMasterEq}) with jump operator 
	\be\label{eq:jumpOp}
	J = \sqrt{\g_1} \sm_1 + \sqrt{\g_2} \sm_2 + \beta
	\ee
	and $H_0 = H_\text{sys} + H_c$ with
	\be
	H_c= \frac{-i}{2} \br{\sqrt{\g_1 \g_2} \spl_2 \sm_1 + \p{\sqrt{\g_1} \spl_1 + \sqrt{\g_2} \spl_2} \beta} + \textrm{Hc}
	\ee
	describing the coupling between the systems.
	
	Now the connection to the quantum trajectory method can be made. 
	There is smooth evolution governed by a Schr\"odinger-like equation with a non-Hermitian effective Hamiltonian 
	\be
	\Heff = H_0 + H',
	\ee
	where $H'=-i J^\dagger J/2$ is the non-Hermitian piece. Additionally, random quantum jumps occur, with probability density $\bra{\psi} \hc{J}J \ket{\psi}$, leading to wavefunction decay \cite{gardiner1992wave}. 
	[This quantum trajectory formalism can be shown to be equivalent to the standard Lindblad master equation dynamics for the density matrix provided we average $\kb{\psi}{\psi}$ over an ensemble of $\ket{\psi}$'s random trajectories \cite{molmer93}.]
	Writing out the effective Hamiltonian explicitly, we find 
	\begin{align}\label{eq:loneHeff}
	\Heff &= H_\text{sys}  
	- \frac{i}{2} \big[ \g_1 \spl_1\sm_1  + \g_2 \spl_2\sm_2 
	+ 2 \sqrt{\g_1 \g_2} \sm_1 \spl_2 \nonumber \\
	&\hspace{11.5mm}+ 2 \beta \p{\sqrt{\g_1} \spl_1 + \sqrt{\g_2} \spl_2} + |\beta|^2 \big],
	\end{align}
	where the $\sm_1 \spl_2$ term survives, not its Hermitian conjugate. This describes an excitation in the first system transferring to the second, but not the other way around.

	\subsection{With transformation}\phantomsection\label{app:transInterpretation}
	By implementing our transformation, the output of system 1 transforms according to Eq.~(\ref{eq:genSys1Field}) in the Heisenberg picture. 
	Thus, the transformed EOM for $s_2$ depends on system 1 and field operators at time $\tw \equiv f(t - \tau)$ and on system 2 operators at the same time $t$. 
	The transformed EOM for $s_1$ only depends on operators at a single time $f(t)$, which can be shifted to $\tw = f(t - \tau) $. 
	Specifically, we start with the transformed Heisenberg EOMs for system operators $s_j$, which are 
	given by Eq.~(\ref{eq:genEOM}) with the replacements 
	$\sm_1(t) \ra \chi(t - \tau) \sm_1(\tw)$ 
	for system 1 evolution including via $c_1$ and 
	$\sm_1(t - \tau) \ra \chi(t - \tau) \sm_1(\tw)$
	for system 2 evolution via $c_2$, and likewise for $\bin$. 
	
	Then we go to the Schr\"odinger evolution of the total transformed state of the systems and bath, $\tilde{\rho}_S(t)$, in the same manner as in App.~\hyperref[appA]{I\,A}. 
	Tracing over the bath yields the reduced density operator describing the composite state of the systems that would produce the transformed wave packet 
	\be
	\rho(t) = \Tr_\textrm{B}(\tilde{\rho}_S(t))
	= \rho_1(\tw)\otimes \rho_2(t).
	\ee 
	This transformed state satisfies the same Lindblad master equation as before with the transformation being accounted for by the change in the time argument of system 1, and the replacements $\sm_1 \ra \chi(t - \tau) \sm_1$ and $\beta \ra \chi(t - \tau) \beta$, that is, the transformed operators acquire an explicit time-dependent prefactor in the Schr\"odinger  picture.
	Here $\rho_1(\tw)$ describes the real system 1 during Stages 1 and 4, the lack of system 1 during Stage 2, and 
	a fictitious system $\tilde{1}$ at time $\xi(T-t + \tau)$ during Stage 3. 
	
	This Heisenberg $\leftrightarrow$ Schr\"odinger conversion works in a similar way as before because the unitary transformation to the field has the effect of transforming system 1 and field operator arguments from $t-\tau$ to $f(t-\tau)$. In the Schr\"odinger picture the same unitary is applied to the state, with the same effect on its argument $t$. Specifically, during the transformation, $\rho_1$'s time argument has slope $-\xi$ corresponding to the fictitious system $\tilde{1}$ evolving backward in time (as the slope is negative) at a new  decay rate due to the $\xi$ scaling. 
	Meanwhile system 2 is continually described by $\rho_2(t)$, as in Eq.~(\ref{eq:compState}).
		
\begin{figure}[h]
	\includegraphics[width=\linewidth]{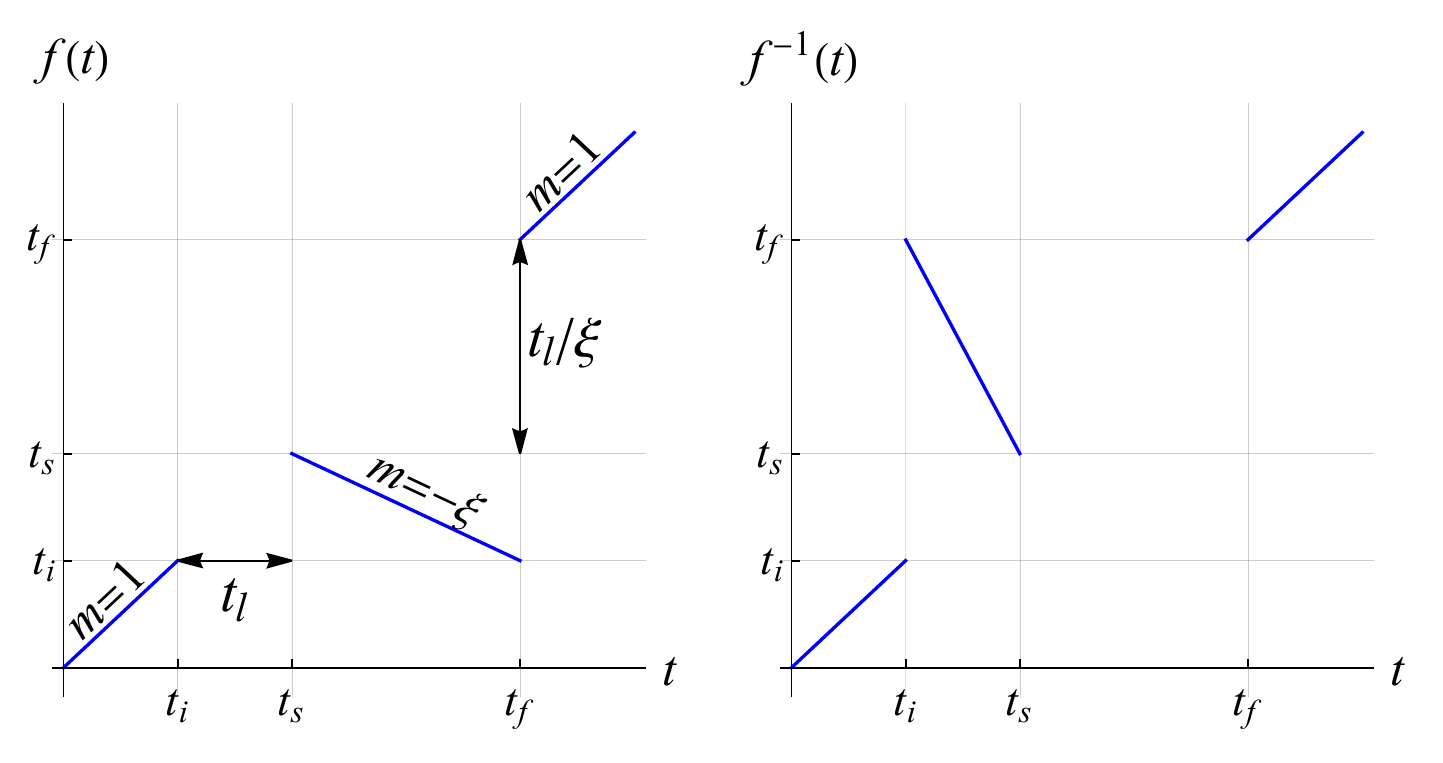}
	\caption{Plots of $f(t)$ and $f^{-1}(t)$ for $X=0$, $\xi = 1/2$.
		 $f^{-1}(t)$ is a fictitious time for system 2 as a function of the time for system 1. In the first plot $m$ denotes the slope of different portions of $f(t)$. The horizontal gap of width $t_l$ is due to the buffering of Stage 2 and the vertical gap of height $t_l/\xi$ is due to us blocking the initial wave packet during Stage 3. 
	}
	\label{fig:fandfInv}
	\end{figure}

	Naively, the jumps in $f(t)$, and hence $\tw$ (see Fig.~\ref{fig:fandfInv}), seem to indicate that system 2 operators would have discontinuities in their evolution as they are being driven by system $\tilde{1}$, which experiences discontinuities when the Stage changes. 
	Yet at $t_i$ system 1 is just leaving one of its ground states (the cavity ground state in Sec.~\ref{sec:example}) and at $t_s$ the decay of the first system is nearly complete (assuming $t_l$ and $T$ are chosen appropriately), so again system 1 will be near a ground state. 
	Hence the coupling between the systems is effectively zero in both cases as system 1 is (at least nearly) in a ground state, which yields zero if acted on by the annihilation operator $\sigma_1^{-}$. This likewise applies to the vertical jump in $f(t)$ at $t_f$. 
	Moreover, system $\tilde{1}$'s evolution will be smoothed out by the gradual switching to and from the vacuum field $V(x,t)$ during Stage 2 (see footnote 3). Thus, system 2 will evolve smoothly even though $f(t)$ has discontinuities.

	
	\section*{Appendix II: Unitary Implementation}\phantomsection\label{appB}
	\renewcommand{\theequation}{II\arabic{equation}}
	\setcounter{equation}{0}
	Here we describe how our simple model for the unitary transformation device is related to a physical implementation proposed by Ref.~\citenum{timereversal} that uses sum-frequency generation (SFG) in a nonlinear medium of length $L$. 
	Ref.~\citenum{timereversal} considers a transformation that 
	takes in two signals, 1 and 2, with frequencies $\om_1$ and $\om_2$,\footnote{Ref.~\citenum{timereversal} uses the notation $\om_1 = \om_s$, $\om_2 = \om_r$, and $\xi = M  = -1/m  > 0$.} 
	that are initially a temporal mode in an arbitrary state and a vacuum state, respectively. These signals are connected using SFG driven by a short classical pump pulse of frequency $\om_p = |\om_2 - \om_1|$ that mixes the signals in the medium. 
	They derive conditions for the output of the transformation to be signal 1 in the vacuum state and signal 2 in a temporal mode that is the time reversed and stretched by $\xi$ counterpart of the initial signal 1 yet is still centered at $\om_2$.
	These conditions include that the group slownesses $\beta'_\eta = 1/v_\eta$ ($\eta = 1,2,p$), which are the inverse group velocities in the medium, are ordered as $\beta'_1 > \beta'_p > \beta'_2$ (for $\omega_2 > \omega_1$, otherwise it is reversed),  as well as things such as phase matching. They verify that these conditions can occur for realistic media. 
	Importantly for QST, they show that the entanglement between the transformed field (which here is the output from system 1) and the pump pulse (and hence the unitary transformation device) stays arbitrarily small in the limit of a classical (i.e.,~strong and coherent) pulse.
	Note that their signals 1 and 2 are our input and output to the unitary transformation device, respectively.
	
	Our Stage 2 and 3 are captured by their transformation. Stage 2 begins at $t_i$ when signal 1, the input field, first enters the transformation device (nonlinear medium). Signal 1 is followed by signal 2 and the pump pulse, which enter the medium at the same time. 
	A three-wave-mixing process (SFG) then begins, and signal 2 (the transformed output) exits the medium, followed by signal 1 and the pump pulse at the same time.
	Using these conditions it follows that to transform a duration $t_l = l/c$ of the input pulse we need a medium of minimum length 
	\be
	L= t_l (\beta'_1 + \beta'_2 - \beta'_p - 1/c)^{-1} 
	> \frac{l}{c \beta'_1 - 1}.
	\ee
	Let $t_\textrm{exit}$ be the time that signal 2 begins exiting the medium at $x=X+L$. Then $t_s = t_\textrm{exit} - L/c$  is the effective time that the transformed output begins to be produced at $x=X$. Thus, the buffering of Stage 2 can physically be interpreted as the effective time between signal 1 entering the medium at $x=X$ and when the transformed signal 2 
	would have been at $x=X$ had it been traveling at the speed of light rather than $\beta'_2$ (i.e., what happens in our simplified device).
	
	
	\bibliographystyle{apsrev4-1}
	\bibliography{draft2}
	
\end{document}